\documentstyle[12pt]{article}

\title{Critical Temperature for the $\lambda (\phi^{4})_{4}$ Within the
$\delta$-Expansion}

\author{Rudnei O. Ramos\\
{\it Dartmouth College, Dept. of Physics and Astronomy,}\\
{\it 6127 Wilder Laboratory,} \\
{\it Hanover, NH 03755, USA} \\
e-mail: rudnei@northstar.dartmouth.edu}

\date{DART-HEP-92/07 \\ October 1992}

\begin{document}

\maketitle

\begin{abstract}

We apply the $\delta$-expansion perturbation scheme to the $\lambda \phi^{4}$
self-interacting scalar field theory in 3+1 D at finite temperature. In the
$\delta$-expansion the interaction term is written as $\lambda (\phi^{2})^{
1 + \delta}$ and $\delta$ is considered as the perturbation parameter. We
compute, in this perturbation approach, the renormalized mass at finite
temperature from which we get the critical temperature. The results are
compared with the usual loop-expansion at finite temperature.

\vspace{0.7 cm}
\noindent
PACS number(s): 11.15.Tk, 05.70.Fh, 64.60.-i .

\end{abstract}

\newpage

Recently a new pertubation scheme in field theory was proposed known as the
\break $\delta$-expansion [\ref{r:1} - \ref{r:4}]. In this novel perturbation
scheme, instead of using Lagrangian parameters for the expansion, like an
expansion in the interaction coupling constant $\lambda$ in the $\lambda
\phi^{4}$ theory (regarding $\lambda$ as a weak-coupling constant) or the
usual loop-expansion (in powers of $\hbar$) the $\delta$-expansion makes use
of an artificial parameter ($ \delta$).

In the usual $\lambda \phi^{4}$ theory in 3+1D, the interaction term is
rewritten as $\lambda M^{4} (M^{-2} \phi^{2})^{1 + \delta}$, where $M$ is an
arbitrary mass parameter introduced to make the coupling
constant $\lambda$ dimensionless. $\delta$ is regarded as a small positive
parameter that can be used as a perturbative parameter in the theory, for
example when Green's functions are computed. Thus the $\lambda \phi^{4}$
Lagrangian for a scalar field $\phi$, in 3+1D, given by

\begin{equation}
{\cal L} = \frac{1}{2} (\partial_{\mu} \phi)^{2} - \frac{\mu^{2}}{2} \phi^{2}
- \frac{\lambda}{4 !} \phi^{4}
\end{equation}

\noindent
is rewritten as

\begin{equation}
{\cal L} = \frac{1}{2} (\partial_{\mu} \phi)^{2} - \frac{\mu^{2}}{2} \phi^{2}
- \frac{\lambda}{4 !} M^{4} (M^{-2} \phi^{2})^{1 + \delta} \: .
\end{equation}

\noindent
Expanding (2) in powers of $\delta$ we get

\begin{equation}
{\cal L} = \frac{1}{2} (\partial_{\mu} \phi)^{2} - \frac{1}{2} ( \mu^{2} +
\frac{\lambda}{4 !} 2 M^{2} ) \phi^{2} - \frac{\lambda}{4 !} M^{2}
\phi^{2} \sum_{n=1}^{\infty} \frac{\delta^{n}}{n !} \left[ \ln (M^{-2}
\phi^{2}) \right]^{n} \: .
\end{equation}

\noindent
If one uses that

\begin{equation}
\sum_{n=0}^{\infty} \frac{\delta^{n}}{n !} \left[ \ln (M^{-2} \phi^{2})
\right]^{n} = \sum_{n=0}^{\infty} \frac{\delta^{n}}{n !}
\frac{ d^{n}}{dk^{n}} (M^{-2} \phi^{2})^{k} |_{k=0} =
e^{\delta \partial_{k}} (M^{-2} \phi^{2})^{k} |_{k=0} \:,
\end{equation}

\noindent
then (3) can be written as[\ref{r:4}]

\begin{equation}
{\cal L} = \frac{1}{2} (\partial_{\mu} \phi)^{2} - \frac{1}{2} ( \mu^{2} +
\frac{\lambda}{4 !} 2 M^{2} ) \phi^{2} - D_{k} \phi^{2k+2} |_{k=0} \:,
\end{equation}

\noindent
where, from the relation (4), $D_{k}$ is a derivative operator given by

\begin{equation}
D_{k} = \frac{\lambda}{4 !} M^{2} \left( e^{\delta \partial_{k}} - 1
\right) \left( M^{-2} \right)^{k} \: .
\end{equation}

In ref. [\ref{r:4}] it was shown that the $n$-point Green's function $G^{(n)}
( x_{1}, x_{2}, \ldots , x_{n})$ can be written as\footnote{In refs.
[\ref{r:1}]
and [\ref{r:2}] the Green's functions are defined on a different way but
the final results are completely analogous.}

\begin{eqnarray}
& G^{(n)} & ( x_{1}, x_{2}, \ldots , x_{n})  =
\prod_{m=0}^{\infty} \frac{1}{m !} \int d^{4}y_{1} d^{4} y_{2} \ldots
d^{4} y_{m} \langle 0 | T \phi(x_{1}) \phi(x_{2}) \ldots \phi(x_{n})
\times \nonumber \\
& D_{k_{1}}& D_{k_{2}} \ldots D_{k_{m}} \left[ \phi^{2}(y_{1})
\right]^{k_{1}+1} \left[ \phi^{2}(y_{2})\right]^{k_{2}+1} \ldots
\left[ \phi^{2}(y_{m})\right]^{k_{m}+1} | 0 \rangle_{c} |_{k=0} \: ,
\end{eqnarray}

\noindent
which can be computed, as shown in ref. [\ref{r:4}], by first considering
the $k_{i}$'s as integers with the same value such that we can draw all
diagrams coming from (7). From (6), if the $k$'s are integers then $D_{k}$ can
be regarded as small and (7) can be computed by ordinary diagrammatic
pertubation. At the end, considering the k's as continuous with $k_{i} \neq
k_{j}, \: i \neq j$, we apply the derivative operators $D_{k_{i}}$ and finally
we make all $k$'s igual to zero.

Once we know how to compute the Green's functions, we can obtain the
renormalized mass $m_{R}$, the renormalized coupling constant $\lambda_{R}$ and
the wave-function renormalization contant $Z$ from the usual definitions:

\begin{equation}
m_{R}^{2} = Z \left[ G_{c}^{(2)} (p^{2}) \right]^{-1} |_{p^{2}=0} \: ,
\end{equation}

\begin{equation}
\lambda_{R} = - Z^{2} G_{c}^{(4)} (0,0,0,0)  \: ,
\end{equation}

\noindent
and

\begin{equation}
Z^{-1} = 1 + \frac{d}{d p^{2}} \left[ G_{c}^{(2)} (p^{2}) \right]^{-1}
|_{p^{2}=0} \: ,
\end{equation}

\noindent
where $G_{c}^{(2)} (p^{2})$ and $G_{c}^{(4)}(p_{1}, p_{2}, p_{3}, p_{4})$
are the connected two-point and four-point Euclidean Green's functions, in
momentum space, respectively.

In refs. [\ref{r:3}] and [\ref{r:4}] $m_{R}^{2}$, $\lambda_{R}$ and $Z$ were
computed for the $\lambda \phi^{4}$ model at arbitrary space-time
dimension $D$,
up to $\delta^{2}$-order.

In this brief report we compute the renormalized mass, given by Eq. (8), at
finite temperature and get the critical temperature for the model of
Lagrangian density given in (1) with mass parameter $\mu^{2} < 0$. It will be
interesting to compare the results with the usual loop-expansion, where for
a Lagrangian density given by (1), we get the well known results (see,
for example, ref. [\ref{r:5}]) for the
renormalized mass, at high temperatures and lowest order in $\lambda$,
$m_{R}^{2}(T)= \mu^{2} + \frac{\lambda}{24} T^{2}$ and the corresponding
critical temperature $T_{c}^{2} = -24 \frac{ \mu^{2}}{\lambda}$.

At lowest order in the coupling constant $\lambda$ the two-point Green's
function $G^{(2)}$ is given by[\ref{r:4}]

\begin{equation}
G^{(2)} = - D_{k} \frac{ (2k + 2) !}{2^{k} k !} \left[ I(m) \right]^{k} |_{k
= 0} \: ,
\end{equation}

\noindent
where $D_{k}$ is given by (6) and $I(m)$ is the loop integral, that at $T \neq
0$ is given by

\begin{equation}
I(m) = \frac{1}{\beta} \sum_{n=-\infty}^{n= + \infty} \int \frac{ d^{3} q}{
(2 \pi)^{3}} \frac{1}{\omega_{n}^{2} + q^{2} + m^{2}} \: ,
\end{equation}

\noindent
and, from (3), $m^{2} = \mu^{2} + \frac{\lambda}{4 !} 2 M^{2}$.

Subtracting the zero temperature divergent contribution\footnote{In
matter of fact we should define how to proceed with the renormalization of
(11), however at first order in the coupling contant
it is easy to show that, from the expressions we will obtain, a
renormalization by the introduction of counter terms in the Lagrangian
density is enough. Thus we are going to refer only to the temperature finite
terms.} of (12), one can write the following expansion [\ref{r:6}]
for $I(m)$ in powers of $m^{2} \beta^{2}$:

\begin{equation}
I(m) = \frac{T^{2}}{12} - \frac{m T}{4 \pi} - \frac{m^{2}}{8 \pi^{2}}
\left( \ln ( \frac{m}{4 \pi T}) + \gamma - \frac{1}{2} \right) +
\frac{m^{4}}{T^{2}} \frac{ \xi(3)}{2^{7} \pi^{4}} + {\cal O} \left(
\frac{m^{6}}{T^{4}} \right) \: .
\end{equation}

Substituting (6) in (11) and using that $e^{\delta \partial_{k}} - 1 =
\delta \frac{\partial}{\partial k} + \frac{\delta^{2}}{2 !} \frac{
\partial^{2}}{\partial k^{2}} + \ldots$  we can compute the renormalized
mass $m_{R}^{2}$, defined by (8), up to third order
in $\delta$ and the corresponding value of the critical temperature at each
order in the expansion in $\delta$, making $\delta = 1$ at the end.

Up to third order in $\delta$ we get the following expression for the
two-point Green's function in lowest order in $\lambda$:

\begin{eqnarray}
& G_{(\delta^{3})}^{(2)} & = - \frac{\lambda}{4 !} 2 M^{2} \left\{
\delta \left[ \ln \left( \frac{I(m)}{2 M^{2}} \right) + 2 \psi(3) - \psi(1)
\right] + \frac{\delta^{2}}{2 !} \left[ \left[ \ln \left( \frac{I(m)}{2 M^{2}}
\right) + 2 \psi(3) - \psi(1)
\right]^{2}  \right. \right. \nonumber \\
&+& \left. \left. 4 \psi'(3) - \psi'(1) \Bigr ] \right.  +
\frac{\delta^{3}}{3 !} \left[ \left[ \ln \left( \frac{I(m)}{2 M^{2}} \right)
+ 2 \psi(3) - \psi(1)
\right]^{3} + 3 \left[ \ln \left( \frac{I(m)}{2 M^{2}} \right) +
2 \psi(3) - \psi(1)
\right] \right. \right.  \nonumber \\
& \times & \left. \left. ( 4 \psi'(3) - \psi'(1)) + 8 \psi''(3) - \psi''(1)
\Bigr ]  \Bigr \} \right. \right.  \: ,
\end{eqnarray}

\noindent
where $\psi (x)$ is the psi-function, $\psi^{(n)}(x) = \frac{d^{n+1}}{
dx^{n+1}} \ln \Gamma(x)$. $I(m)$ is given by (13) and $ m^{2} = \mu^{2} +
\frac{\lambda}{4 !}
2 M^{2}$.

The wave-function renormalization constant $Z$ defined by (10) will depend
on two-point Green's functions of order $\lambda^{2}$ and higher, which
are dependent on external momenta $p_{i}^{2}$. Therefore, from (10)

\begin{equation}
Z= 1 + {\cal O}(\lambda^{2}) \: .
\end{equation}

Substituting (14) and (15) in (8), we get the following expression for the
renormalized mass up to order $\lambda$ in the coupling constant and up
to third order in $\delta$:

\begin{equation}
m_{R}^{2} = \mu^{2} + \frac{\lambda}{4 !} 2 M^{2} - G_{(\delta^{3})}^{(2)}
+ {\cal O}(\lambda^{2})  \: ,
\end{equation}

\noindent
with $G_{(\delta^{3})}^{(2)}$ given by (14).

The whole dependence of (16) on the arbitrary mass parameter $M$
can be removed by requiring that [\ref{r:6}, \ref{r:7}]

\begin{equation}
\frac{\partial m_{R}^{2}}{\partial M^{2}} = 0 \: ,
\end{equation}

\noindent
at each order in the $\delta$-expansion.

{}From the condition (17) we get the following expressions for the mass
parameter $M$ at each order in $\delta$, for $\delta = 1$, and in
lowest order in $\lambda$, at finite temperature:

\begin{equation}
2 M^{2} = \left\{
\begin{array}{ll}
\frac{T^{2}}{12} \exp \left[ 2 \psi(3) - \psi(1) \right] \: , & \mbox{ up to
order $\delta$} \\
\frac{T^{2}}{12} \exp \left[ 2 \psi(3) - \psi(1) - \sqrt{\psi'(1) - 4 \psi'(
3)} \: \: \right] \: , & \mbox{up to order $\delta^{2}$} \\
\frac{T^{2}}{12} \exp \left[ 2 \psi(3) - \psi(1) + y \right] \: ,& \mbox{up to
order $\delta^{3}$}
\end{array}
\right. \: ,
\end{equation}

\noindent
where $y= \left[ 4 \psi''(3) - \frac{1}{2}
\psi''(1) - \sqrt{D} \: \right]^{\frac{1}{3}} + \left[ 4 \psi''(3) -
\frac{1}{2}
\psi''(1) + \sqrt{D} \: \right]^{\frac{1}{3}}$ with \break $D= \left[ 4
\psi''(3) -
\frac{1}{2} \psi''(1) \right]^{2} + \left[ 4 \psi'(3) - \psi'(1)
\right]^{3}$, in the above equation.

Using (18) for $M^{2}$ back on (16), we get the following expression for
the renormalized mass at finite temperature, up to orders $\delta$,
$\delta^{2}$ and $\delta^{3}$, respectively:

\begin{equation}
m_{R}^{2}(T) = \left\{
\begin{array}{l}
\mu^{2} + \frac{\lambda}{4 !} \frac{T^{2}}{12} \exp \left[ 2 \psi(3) -
\psi(1) \right] \: ,  \\
\mu^{2} + \frac{\lambda}{4 !} \frac{T^{2}}{12} \exp \left[ 2 \psi(3) -
\psi(1) - \sqrt{\psi'(1) - 4 \psi'(3)} \: \: \right] \left(1 +
\sqrt{\psi'(1) -
4 \psi'(3)} \:\: \right) \: ,  \\
\mu^{2} + \frac{\lambda}{4 !} \frac{T^{2}}{12} \exp \left[ 2 \psi(3) -
\psi(1) + y \right] \left[ 1 - y - \frac{y^{2}}{2} + \frac{1}{2}
\left( 4 \psi'(3) - \psi'(1) \right) \right] \: .

\end{array}
\right.
\end{equation}

If one uses the numerical values for the psi-functions [\ref{r:8}] in
(19), we get the following values for the critical temperature:
$T_{c_{\delta}}^{2} \simeq -25.5 \frac{\mu^{2}}{\lambda}$,
$T_{c_{\delta^{2}}}^{
2} \simeq - 26.3 \frac{\mu^{2}}{\lambda}$ and $T_{c_{\delta^{3}}}^{2}
\simeq - 18.0 \frac{\mu^{2}}{\lambda}$, from $m_{R}^{2}(T)$ up to orders
$\delta$, $\delta^{2}$ and $\delta^{3}$, respectively. Although
apparently these results seem to diverge from the usual value of the critical
temperature at lowest order in $\lambda$, $T_{c}^{2}= - 24 \frac{\mu^{2}}{
\lambda}$, it is easy to show that the expansion (16) is the expansion in
powers of $\delta$ of

\begin{equation}
m_{R}^{2} = \mu^{2} + \frac{\lambda}{4 !} M^{2} \frac{ (2 \delta + 2) !}
{2^{\delta} \delta !} \left[ M^{-2} I(m) \right]^{\delta} \: ,
\end{equation}

\noindent
such that, in the high temperature limit, for $I(m) \simeq \frac{T^{2}}{12}$
and for $\delta =1$, we get the usual result, $T_{c}^{2} = - 24 \frac{
\mu^{2}}{\lambda}$, for the critical temperature, from (20). However it is
remarkable that even for $\delta = 1$ and at lowest order, the expansion (16)
produces a value for the critical temperature consistent with the result
obtained at lowest order in
the coupling constant.
However when we go beyond of order $\delta$,
we should have to consider terms of higher order in the coupling
constant; for the Green's functions expressed in
terms of Feynman diagrams, we should have to compute diagrams with more
than one vertex (each vertex implies in
an extra derivative operator, given by Eq. (6)). Hence our results are
strictly correct only to order $\delta$.

Since the $\delta$-expansion provides us with one extra parameter to expand the
theory (the $\delta$ itself) and at the same time generates a mass term given
by $ \frac{\lambda}{4 !} 2 M^{2}$, at finite temperatures this generates a
further improvement in relation to the usual expansion in loops.
While in the loop expansion we
have to deal with the usual problems of infrared divergences arriving from
small masses, at temperatures close to $T_{c}$, in the $\delta$-expansion
we get a quite
different behavior at $T \sim T_{c}$ due to the introduction of the extra term
$ \frac{\lambda}{4 !} 2 M^{2}$ in the propagator of the scalar field. At
lowest order in the $\delta$-expansion (and also up to order $\delta^{2}$),
from the results above, we obtained a critical temperature that is slight
larger than the usual result, therefore the effective mass of the
scalar field is comparetively smaller than that of the loop-expansion.
As we can express the correlation length $\xi(T)$ as $m^{-1}(T)$, we get then
that $\xi_{\delta-expansion} > \xi_{loop-expansion}$, suggesting that
we are incorporating some infrared corrections in the theory by using the
$\delta$-expansion. In this way the $\delta$-expansion is reminiscent of the
$\epsilon$-expansion in statistical mechanics [\ref{r:9}].
Work exploring these
characteristics of the $\delta$-expansion is currently in progress, with the
hope that infrared divergences in the electroweak transition can be studied
in the near future.

\vspace{1.0cm}
The author thanks M. Gleiser for discussions and Conselho Nacional de
Desenvolvimento Cient\'{\i}fico e Tecnol\'ogico - CNPq (Brazil) for
financial support.

\end{document}